\documentclass[twocolumn,english,onecolumn]{IEEEtran}
\usepackage[T1]{fontenc}
\usepackage{babel}
\usepackage{array}
\usepackage{prettyref}
\usepackage{multirow}
\usepackage{amsmath}
\usepackage{amssymb}
\usepackage{graphicx}
\PassOptionsToPackage{normalem}{ulem}
\usepackage{ulem}
\usepackage[unicode=true,
 bookmarks=true,bookmarksnumbered=true,bookmarksopen=true,bookmarksopenlevel=1,
 breaklinks=false,pdfborder={0 0 0},backref=false,colorlinks=false]
 {hyperref}
\hypersetup{pdftitle={Your Title},
 pdfauthor={Your Name},
 pdfpagelayout=OneColumn, pdfnewwindow=true, pdfstartview=XYZ, plainpages=false}
\usepackage{breakurl}

\makeatletter

\providecommand{\tabularnewline}{\\}

\usepackage{babel}
\ifCLASSOPTIONcompsoc
\else
\fi
\usepackage{cite}\usepackage{MnSymbol}\usepackage{subfigure}

\usepackage{balance}

\let\labelindent\relax
\usepackage{enumitem}

\usepackage{babel}

\makeatother

\begin{document}

\title{Failure Mitigation in Linear, Sesquilinear and Bijective Operations
On Integer Data Streams Via Numerical Entanglement}

\author{Mohammad Ashraful Anam and~Yiannis Andreopoulos$^{*}$%
\thanks{\protect\textsuperscript{{*}}Corresponding author. The authors are
with the Electronic and Electrical Engineering Department, University
College London, Roberts Building, Torrington Place, London, WC1E 7JE,
Tel. +44 20 7679 7303, Fax. +44 20 7388 9325 (both authors), Email:
\{mohammad.anam.10, i.andreopoulos\}@ucl.ac.uk. This work appeared
in the 21st IEEE International On-Line Testing Symposium (IOLTS 2015)
and was supported by EPSRC, project EP/M00113X/1.%
}}
\maketitle
\begin{abstract}
A new \textit{roll-forward} technique is proposed that recovers from
any single \textit{fail-stop} failure in $M$ integer data streams
($M\geq3$) when undergoing linear, sesquilinear or bijective (LSB)
operations, such as: scaling, additions/subtractions, inner or outer
vector products and permutations. In the proposed approach, the $M$
input integer data streams are linearly superimposed to form $M$
\emph{numerically entangled} integer data streams that are stored
in-place of the original inputs. A series of LSB operations can then
be performed directly using these entangled data streams. The output
results can be extracted from any $M-1$ entangled output streams
by additions and arithmetic shifts, thereby guaranteeing robustness
to a fail-stop failure in any single stream computation. Importantly,
unlike other methods, the number of operations required for the entanglement,
extraction and recovery of the results is linearly related to the
number of the inputs and does not depend on the complexity of the
performed LSB operations. We have validated our proposal in an Intel
processor (Haswell architecture with AVX2 support) via convolution
operations. Our analysis and experiments reveal that the proposed
approach incurs only $1.8\%$ to $2.8\%$ reduction in processing
throughput in comparison to the failure-intolerant approach. This
overhead is 9 to 14 times smaller than that of the equivalent checksum-based
method. Thus, our proposal can be used in distributed systems and
unreliable processor hardware, or safety-critical applications, where
robustness against fail-stop failures becomes a necessity. \end{abstract}


\section{Introduction}

\IEEEPARstart{T}{he increase} of integration density \cite{nicolaidis2012design}
and aggressive voltage/frequency scaling in processor and custom-hardware
designs \cite{alameldeen2011energy}, along with the ever-increasing
tendency to use commercial off-the-shelf processors to create vast
computing clusters, have decreased the mean-time-to-failure of modern
computing systems. Therefore, it is now becoming imperative for distributed
computing systems to provide for fail-stop failure mitigation \cite{gotoda2012task},
i.e., recover from cases where one of their processor cores becomes
unresponsive or does not return the results within a predetermined
deadline. Applications that are particularly prone to fail-stop failures
include distributed systems like grid computing \cite{foster2008cloud},
sensor-network \cite{kurschl2009combining}, webpage, or multimedia
retrieval and object or face recognition in images \cite{yang2004two},
financial computing \cite{peng2010parallel}, etc. The compute- and
memory-intensive parts of these applications comprise linear, sesquilinear
(also known as ``one-and-half linear'') and bijective operations,
collectively called LSB operations in this paper. These operations
are typically performed using single or double-precision floating-point
inputs or, for systems requiring exact reproducibility and/or reduced
hardware complexity, 32-bit or 64-bit integer or fixed-point inputs.
Thus, ensuring robust recovery from fail-stop failures for applications
comprising integer LSB operations is of paramount importance.

\subsection{Summary of Prior Work }

Existing techniques that can ensure recovery from fail-stop failures
comprise two categories: \emph{(i)} \textit{roll-back} via checkpointing
and recomputation \cite{ren2007failure,chen2005fault}, i.e., methods
that periodically save the state of all running processes, such that
the execution can be rolled back to a ``safe state'' in case of
failures;\emph{ (ii)} \textit{roll-forward} methods producing additional
``checksum'' inputs/outputs \cite{chen2005fault,stefanidis2004weighted,chen2009optimal}
such that the missing results from a core failure can be recovered
from the remaining cores without recomputation. Examples of roll-forward
methods include algorithm-based fault-tolerance (ABFT) and modular
redundancy (MR) methods \cite{huang1984algorithm,chen2005fault,luk1985weighted,stefanidis2004weighted,sloan2012algorithmic,rexford1992partitioned,chen2009optimal,engelmann2009case}.
Although no recomputation is required in roll-forward methods (thereby
ensuring quick recovery from a failure occurrence), checksum-based
methods can incur significant computational and energy-consumption
overhead because of the additional checksum-generation and redundant
computations required \cite{rizzo1997effective}.

\subsection{Contribution }

We propose a new roll-forward failure-mitigation method for linear,
sesquilinear (also known as one-and-half linear) or bijective operations
performed in integer data streams. Examples of such operations are
element-by-element additions and multiplications, inner and outer
vector products, sum-of-squares and permutation operations. They are
the building blocks of algorithms of foundational importance, such
as: matrix multiplication \cite{goto2008anatomy,huang1984algorithm},
convolution/cross-correlation \cite{anam2012throughput}, template
matching for search algorithms \cite{anastasia2010software}, covariance
calculations \cite{yang2004two}, integer-to-integer transforms \cite{lin2000packed}
and permutation-based encoding systems \cite{fenwick1996burrows},
which form the core of the applications discussed earlier. Because
our method performs linear superpositions of input streams onto each
other, it ``entangles'' input streams together and we term it as
\emph{numerical entanglement}. Our approach guarantees recovery from
any single stream-processing failure without requiring recomputation.
Importantly, numerical entanglement does not generate additional ``checksum''
or duplicate streams and does not depend on the specifics of the LSB
operation performed. It is therefore found to be extremely efficient
in comparison to checksum-based methods that incur overhead proportional
to the complexity of the operation performed.

\subsection{Paper Organization }

In \prettyref{sec:ABFT_MR_vs_Entanglement}, we introduce checksum
based methods and MR for fail-stop failure recovery in numerical stream
processing. In \prettyref{sec:From-Numerical-Packing-to-Entanglement}
we introduce the notion of numerical entanglement and demonstrate
its inherent reliability for LSB processing of integer streams. \prettyref{sec:Linear_processing}
presents the complexity of numerical entanglements within integer
linear or sesquilinear operations. Section \ref{sec:Experiments}
presents experimental comparisons and \prettyref{sec:Conclusions}
presents some concluding remarks.

\section{Checksum/MR-based Methods versus Numerical Entanglement \label{sec:ABFT_MR_vs_Entanglement}}

Consider a series of $M$ input streams of integers, each comprising
$N_{\text{in}}$ samples%
\footnote{Notations: Boldface uppercase and lowercase letters indicate matrices
and vectors, respectively; the corresponding italicized lowercase
indicate their individual elements, e.g. $\mathbf{A}$ and $a_{m,n}$;
$\hat{d}$ denotes the recovered value of $d$ after disentanglement;
all indices are integers. Operators: superscript $\text{T}$ denotes
transposition; $\left\lfloor a\right\rfloor $ is the largest integer
that is smaller or equal to $a$ (floor operation); $\left\lceil a\right\rceil $
is the smallest integer that is larger or equal to $a$ (ceil operation);
$a\ll b$ and $a\gg b$ indicate left and right arithmetic shift of
integer $a$ by $b$ bits with truncation occurring at the most-significant
or least significant bit, respectively; $a\,\text{mod}\, b=a-\left\lfloor \frac{a}{b}\right\rfloor b$
is the modulo operation. %
} ($M\geq3$): 

\begin{equation}
\mathbf{c}_{m}=\begin{bmatrix}c_{m,0} & \ldots & c_{m,N_{\text{in}}-1}\end{bmatrix},\:0\leq m<M.\label{eq:M_input_data_streams}
\end{equation}
These may be the elements of $M$ rows of a matrix of integers, or
a set of $M$ input integer streams of data to be operated upon with
an integer kernel $\mathbf{g}$. This operation is performed by:

\[
\forall m:\;\mathbf{d}_{m}=\mathbf{c}_{m}\:\text{op}\:\mathbf{g}
\]
\begin{equation}
\mathbf{\text{op}\in\left\{ +\mathrm{,}-\mathrm{,}\times\mathrm{,}\left\langle \centerdot,\centerdot\right\rangle ,\otimes\mathrm{,}\left(\begin{array}{c}
\mathfrak{I}\\
\mathfrak{G}
\end{array}\right)\mathrm{,}\star\right\} }\label{eq:operation_on_inputs}
\end{equation}
with $\mathbf{d}_{m}$ the $m$th vector of output results (containing
$N_{\text{out}}$ values) and $\text{op}$ any LSB operator such as
element-by-element addition/subtraction/multiplication, inner/outer
product, permutation%
\footnote{We remark that we consider LSB operations that are \emph{not} data-dependent,
e.g., permutations according to fixed index sets as in the Burrows-Wheeler
transform \cite{fenwick1996burrows}. %
} (i.e., bijective mapping from the sequential index set $\mathbf{\mathfrak{I}}$
to index set $\mathbf{\mathfrak{G}}$ corresponding to $\mathbf{g}$)
and circular convolution or cross-correlation with $\mathbf{g}$.
Beyond the single LSB operator indicated in \eqref{eq:operation_on_inputs},
we can also assume\emph{ series} of such operators applied consecutively
in order to realize higher-level algorithmic processing, e.g., multiple
consecutive additions, subtractions and scaling operations with pre-established
kernels followed by circular convolutions and permutation operations.
Conversely, the input data streams can also be left in their native
state (i.e., stored in memory), if $\text{op}=\left\{ \times\right\} $
and $\mathbf{g}=1$.

\subsection{Checksum-based Methods}

In their original (or ``pure'') form, the input data streams of
\eqref{eq:M_input_data_streams} are uncorrelated and one input or
output element cannot be used for the recovery of another without
inserting some form of coding or redundancy. This is conventionally
achieved via checksum-based methods \cite{huang1984algorithm,chen2005fault,luk1985weighted,stefanidis2004weighted,nair1988linearCode,sloan2012algorithmic,rexford1992partitioned}.
Specifically, one \emph{additional} input stream is created, which
comprises \emph{checksums} of the original inputs:

\begin{equation}
\mathbf{r}=\begin{bmatrix}r_{0} & \ldots & r_{N_{\text{in}}-1}\end{bmatrix},\label{eq:P_redundant_data_streams}
\end{equation}
by using, for example, the sum of groups of $M$ input samples \cite{rexford1992partitioned,sloan2012algorithmic}
at position $n$ in each stream, $0\leq n<N_{\text{in}}$:

\begin{equation}
\forall n:\; r_{n}=\sum_{m=0}^{M-1}c_{m,n}.\label{eq:redundant_element_definition}
\end{equation}
Then the processing is performed in all input streams $\mathbf{c}_{0},\ldots,\mathbf{c}_{M-1}$
and in the checksum input stream $\mathbf{r}$ (each running on a
different core) by:

\begin{equation}
\begin{bmatrix}\mathbf{d}_{0}\\
\vdots\\
\mathbf{d}_{M-1}\\
\mathbf{e}
\end{bmatrix}=\begin{bmatrix}\mathbf{c}_{0}\\
\vdots\\
\mathbf{c}_{M-1}\\
\mathbf{r}
\end{bmatrix}\:\text{op}\:\mathbf{g},\label{eq:operation_on_inputs_and_redundant_inputs}
\end{equation}
Any single fail-stop core failure in the group of $M+1$ cores executing
\eqref{eq:operation_on_inputs_and_redundant_inputs} can be recovered
from using the remaining $M$ output streams. As discussed in partitioning
schemes for checksum-based methods and ABFT \cite{rexford1992partitioned,sloan2012algorithmic},
the recovery capability can be increased by using additional weighted
checksums.

\subsection{Proposed Numerical Entanglement}

Numerical entanglement mixes the inputs prior to processing using
linear superposition, and ensures the results can be recovered via
a mixture of shift-add operations. Specifically, considering $M$
($M\geq3$) input streams $\mathbf{c}_{m}$, $0\leq m<M$ (each comprising
$N_{\text{in}}$ integer samples), each element of the $m$th entangled
stream denoted by $\epsilon_{m,n}$ ($0\leq n<N_{\text{in}}$), comprises
the superposition of two input elements $c_{x,n}$ and $c_{y,n}$
from different input streams $x$ and $y$, i.e., $0\leq x,y<M$ and
$x\neq y.$ The LSB operation $\text{op}$ with kernel $\mathbf{g}$
is carried out with $M$ independent cores utilizing the entangled
input streams directly, thereby producing the entangled output streams
$\boldsymbol{\delta}_{m}$ (each comprising $N_{\text{out}}$ integer
samples). These can be disentangled to recover the final results $\mathbf{\hat{d}}_{m}$.
Any single fail-stop failure in the $M$ processor cores can be recovered
from the results of the remaining $M-1$ cores utilizing additions
and shift operations. 

The complexity of entanglement, disentanglement (extraction) and recovery
does not depend on the complexity of the operator $\text{op}$, or
on the length of the kernel (operand) $\mathbf{g}$. The entangled
inputs can be written in-place and no additional storage or additional
operations are needed during the execution of the actual operation.
The entire process is also suitable for stream processors with entanglement
applied as data within each input stream is being read. Unlike checksum
or MR methods, numerical entanglement does not use additional processor
cores, and the only detriment is that the dynamic range of the entangled
inputs $\boldsymbol{\epsilon}_{m}$ is somewhat increased in comparison
to the original inputs $\mathbf{c}_{m}$. However, as it will be demonstrated
in the next section, this increase depends on the number of jointly-entangled
inputs, $M$, i.e., the desired failure recovery capability. Therefore,
one can be traded for the other.

\section{Numerical Entanglement for Fail-Stop Reliability in LSB Operations
\label{sec:From-Numerical-Packing-to-Entanglement}}

We first illustrate our approach via its simplest instantiation, i.e.,
entanglement of $M=3$ inputs, and then present its general application
and discuss its properties.

\subsection{Numerical Entanglement in Groups of $M=3$ Inputs}

\subsubsection{Entanglement}

\begin{figure}[tbh]
\begin{centering}
\includegraphics[scale=0.40]{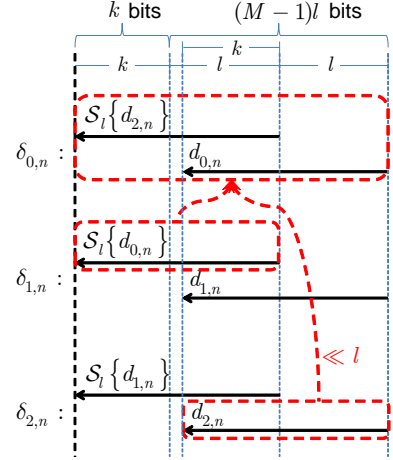}
\par\end{centering}

\protect\caption{Illustration of three entangled outputs after integer LSB processing.
The solid arrows indicate the maximum attainable dynamic range of
each output $d_{0,n}$, $d_{1,n}$ and $d_{2,n}$. The dotted rectangles
and arrows illustrate that the contents of entangled output $\delta_{0,n}$
are contained within the two other entangled outputs. \label{fig:Entanglement-with-M_equal_3}}
\end{figure}

In the simplest form of entanglement ($M=3$), each triplet of input
samples of the three integer streams, $c_{0,n}$, $c_{1,n}$ and $c_{2,n}$,
$0\leq n<N_{\text{in}}$, produces the following entangled triplet
via the superposition operations:

\begin{eqnarray}
\mathit{\epsilon}_{0,n} & = & \mathcal{S}_{l}\left\{ c_{2,n}\right\} +c_{0,n}\nonumber \\
\epsilon_{1,n} & = & \mathcal{S}_{l}\left\{ c_{0,n}\right\} +c_{1,n}\label{eq:entanglement_M_equal_1}\\
\epsilon_{2,n} & = & \mathcal{S}_{l}\left\{ c_{1,n}\right\} +c_{2,n}\nonumber 
\end{eqnarray}
where: 

\begin{equation}
\mathcal{S}_{l}\left\{ c\right\} \equiv\left\{ \begin{array}{c}
\;\;\left(c\ll l\right),\;\;\;\;\,\text{if}\; l\geq0\\
\left[c\gg\left(-l\right)\right],\;\text{if}\; l<0
\end{array}\right.
\end{equation}
is the left or right arithmetic shift of $c$ by $l$ bits. If we
assume that the utilized integer representation comprises $w$ bits,
the $l$-bit left-shift operations of \eqref{eq:entanglement_M_equal_1}
must be upper-bounded by $w$ to avoid overflow. Therefore, if the
dynamic range of the input streams $\mathbf{c}_{0}$, $\mathbf{c}_{1}$,
$\mathbf{c}_{2}$ is $l+k$ bits:

\begin{equation}
2l+k\leq w\label{eq:entanglement-condition-M_eq_1}
\end{equation}
in order to ensure no overflow happens from the arithmetic shifts
of \eqref{eq:entanglement_M_equal_1}. The values for $l$ and $k$
are chosen such that $l+k$ is maximum within the constraint of \eqref{eq:entanglement-condition-M_eq_1}
and $k\leq l$. Via the application of LSB operations, each $\boldsymbol{\epsilon}_{m}$
entangled input stream ($0\leq m<M$) is converted to the entangled
output stream%
\footnote{For the particular cases of: $\text{op}\in\left\{ +,-\right\} $,
$\mathbf{g}$ must also be entangled with itself via: $g_{n}\leftarrow\mathcal{S}_{l}\left\{ g_{n}\right\} +g_{n}$,
in order to retain the homomorphism of the performed operation. %
} $\boldsymbol{\delta}_{m}$ (which contains $N_{\text{out}}$ values): 

\begin{equation}
\forall m:\boldsymbol{\;\delta}_{m}=\left(\boldsymbol{\epsilon}_{m}\:\text{op}\:\mathbf{g}\right).\label{eq:operation_on_inputs-1-1}
\end{equation}
A conceptual illustration of the entangled outputs after \eqref{eq:entanglement_M_equal_1}
and \eqref{eq:operation_on_inputs-1-1} is given in Fig. \ref{fig:Entanglement-with-M_equal_3}.
Our description until this point indicates a key aspect: $l$ bits
of dynamic range are used \emph{within} each entangled input/output
in order to achieve recovery from one fail-stop failure occurring
in the computation of $\delta_{0,n}$, $\delta_{1,n}$ or $\delta_{2,n}$.
As a practical instantiation of \eqref{eq:entanglement_M_equal_1},
we can set $w=32$, $l=11$ and $k=10$ in a signed 32-bit integer
configuration. 

We now describe the disentanglement and recovery process. The reader
can also consult Fig. \ref{fig:Entanglement-with-M_equal_3}.

\subsubsection{Disentanglement and Recovery}

We can disentangle the outputs by ($0\leq n<N_{\text{out}}$): 

\begin{eqnarray}
d_{\textrm{temp}} & = & \delta_{2,n}-\mathcal{S}_{l}\left\{ \delta_{1,n}\right\} \nonumber \\
\hat{d}_{2,n} & = & \mathcal{S}_{-2\left(w-l\right)}\left\{ \mathcal{S}_{2\left(w-l\right)}\left\{ d_{\textrm{temp}}\right\} \right\} \nonumber \\
\hat{d}_{0,n} & = & \mathcal{S}_{-2l}\left\{ -\left(d_{\textrm{temp}}-\hat{d}_{2,n}\right)\right\} \label{eq:integer_disentanglement}\\
\hat{d}_{1,n} & = & \delta_{1,n}-\mathcal{S}_{l}\left\{ \hat{d}_{0,n}\right\} \nonumber 
\end{eqnarray}
The first three parts of \eqref{eq:integer_disentanglement} assume
a $2w$-bit integer representation is used for the interim operations,
as the temporary variable $d_{\textrm{temp}}$ is stored in $2w$-bit
integer representation. However, all recovered outputs, $\hat{d}_{0,n}$,
$\hat{d}_{1,n}$ and $\hat{d}_{2,n}$, require only $w$ bits. 

Explanation of \eqref{eq:integer_disentanglement}---see also Fig.
\ref{fig:Entanglement-with-M_equal_3}: The first part creates a composite
number comprising $\hat{d}_{0,n}$ in the $l+k$ most-significant
bits and $\hat{d}_{2,n}$ in the $2l$ least-significant bits (therefore,
$d_{\textrm{temp}}$ requires $3l+k$ bits). In the second part, $\hat{d}_{2,n}$
is extracted by: \emph{(i)} discarding the $\left(2w-2l\right)$ most-significant
bits; \emph{(ii)} arithmetically shifting the output down to the correct
range. The third part of \eqref{eq:integer_disentanglement} uses
$\hat{d}_{2,n}$ to recover $\hat{d}_{0,n}$ and, in the fourth part
of \eqref{eq:integer_disentanglement}, $\hat{d}_{0,n}$ is used to
recover $\hat{d}_{1,n}$. 

\emph{Remark 1 (operations within $w$ bits):} To facilitate our exposition,
the first three parts of \eqref{eq:integer_disentanglement} are presented
under the assumption of a $2w$-bit integer representation. However,
it is straightforward to implement them via $w$-bit integer operations
by separating $d_{\text{temp}}$ into two parts of $w$ bits and performing
the operations separately within these parts\emph{.}

\emph{Remark 2 (recovery without the use of }$\delta_{0,n}$\emph{):}
Notice that \eqref{eq:integer_disentanglement} does not use $\delta_{0,n}$.
This is a crucial element of our approach: since $\hat{d}_{0,n}$,
$\hat{d}_{1,n}$ and $\hat{d}_{2,n}$ were derived without using $\delta_{0,n}$,
full recovery of all outputs takes place even with the loss of one
entangled stream. We are able to do this because, for every $n$,
$0\leq n<N_{\text{out}}$, $\delta_{1,n}$ and $\delta_{2,n}$ contain
$\hat{d}_{0,n}$ and $\hat{d}_{2,n}$, which suffice to recreate $\delta_{0,n}$
if the latter is not available due to a fail-stop failure. This link
is pictorially illustrated in Fig. \ref{fig:Entanglement-with-M_equal_3}.
Since the entangled pattern is cyclically-symmetric, it is straightforward
to demonstrate that recovery from loss of any single out the three
output streams is possible following the same approach.

\emph{Remark 3 (dynamic range):} Bit $l+k$ within each recovered
output $\hat{d}_{0,n}$, $\hat{d}_{1,n}$ and $\hat{d}_{2,n}$ represents
its sign bit. Given that: \emph{(i)} each entangled output comprises
the addition of two outputs (with one of them left-shifted by $l$
bits); \emph{(ii)} the entangled outputs must not exceed $2l+k$ bits,
we deduce that the outputs of the LSB operations must not exceed the
range 

\begin{equation}
\forall n:\; d_{0,n},d_{1,n},d_{2,n}\in\left\{ -\left(2^{l+k-1}-2^{l}\right),2^{l+k-1}-2^{l}\right\} .\label{eq:dynamic_range}
\end{equation}
Therefore, \eqref{eq:dynamic_range} comprises the range permissible
for the LSB operations of \eqref{eq:operation_on_inputs-1-1} with
the entangled representation of \eqref{eq:entanglement_M_equal_1}.
Thus, we conclude that, for integer outputs produced by the LSB operations
of \eqref{eq:operation_on_inputs-1-1} with range bounded by \eqref{eq:dynamic_range},
the extraction mechanism of \eqref{eq:integer_disentanglement} is
\emph{necessary and sufficient} for the recovery of \emph{any single
stream in} $\delta_{0,n}$, $\delta_{1,n}$, $\delta_{2,n}$ for all
stream positions $n$, $0\leq n<N_{\text{out}}$.

\subsection{Generalized Entanglement in Groups of $M$ Inputs ($M\geq3$)}

We extend the proposed entanglement process to using $M$ inputs and
providing $M$ entangled descriptions, each comprising the linear
superposition of two inputs. This ensures that, for every\emph{ }$n$
($0\leq n<N_{\text{out}}$)\emph{, any }single failure will be recoverable
within each group of $M$ output samples. 

The condition for ensuring that overflow is avoided is

\begin{equation}
\left(M-1\right)l+k\leq w\label{eq:M_l_k_constraint}
\end{equation}
and the dynamic range supported for all outputs is ($\forall m,n$): 

\begin{equation}
d_{m,n}\in\left\{ -2^{\left(M-3\right)l+k}\left(2^{l-1}-1\right),2^{\left(M-3\right)l+k}\left(2^{l-1}-1\right)\right\} \label{eq:M_dynamic_range}
\end{equation}
We now define the following operator that generalizes the proposed
numerical entanglement process:

\begin{equation}
\boldsymbol{\mathcal{E}}=\begin{bmatrix}1 & 0 & \cdots & 0 & \mathcal{S}_{l}\\
\mathcal{S}_{l} & 1 & \cdots & 0 & 0\\
 &  & \ddots\\
0 & \cdots & \mathcal{S}_{l} & 1 & 0\\
0 & \cdots & 0 & \mathcal{S}_{l} & 1
\end{bmatrix}_{M\times M}\label{eq:entangle_matrix}
\end{equation}
with $\boldsymbol{\mathcal{E}}$ the circulant matrix operator comprising
cyclic permutations of the $1\times M$ vector $\begin{bmatrix}1 & 0 & \cdots & 0 & \mathcal{S}_{l}\end{bmatrix}$. 

As before, in the generalized entanglement in groups of $M$ streams,
the values for $l$ and $k$ are chosen such that $l+k$ is maximum
within the constraint of \eqref{eq:M_l_k_constraint} and $k\leq l$.
Moreover, the exact same principle applies, i.e., pairs of inputs
are entangled together (with one of the two shifted by $l$ bits)
to create each entangled input stream of data. Any LSB operation is
then performed directly on these input streams and\emph{ any }single
fail-stop failure will be recoverable within each group of $M$ outputs.
For every input stream position $n$, $0\leq n<N_{\text{in}}$, the
entanglement vector performing the linear superposition of pairs out
of $M$ inputs is now formed by:

\begin{equation}
\begin{bmatrix}\epsilon_{0,n} & \cdots & \epsilon_{M-1,n}\end{bmatrix}^{\text{T}}=\mathcal{\boldsymbol{\mathcal{E}}}\left\{ \begin{bmatrix}c_{0,n} & \cdots & c_{M-1,n}\end{bmatrix}^{\text{T}}\right\} .\label{eq:entanglement_M_general}
\end{equation}

After the application of \eqref{eq:operation_on_inputs-1-1}, we can
disentangle every output stream element $\delta_{m,n}$, $0\leq n<N_{\text{out}}$,
as follows. We first identify the unavailable entangled output stream
$\boldsymbol{\delta}_{r}$ (with $0\leq r<M$) due the single core
failure. Then, we produce the $2w$-bit temporary variable $d_{\textrm{temp}}$
by:

\begin{equation}
d_{\textrm{temp}}=\sum_{m=0}^{M-2}\left(-1\right)^{m}\mathcal{S}_{(M-2-m)l}\left\{ \delta_{\left(r+1+m\right)\text{mod}M,n}\right\} .\label{eq:integer_disentanglement_partial_M_general}
\end{equation}
Notice that \eqref{eq:integer_disentanglement_partial_M_general}
does not use $\boldsymbol{\delta}_{r}$. We can then extract the value
of $\hat{d}_{r,n}$ and $\hat{d}_{\left(r+M-1\right)\text{mod}M,n}$
directly from $d_{\textrm{temp}}$:

\begin{eqnarray}
\hat{d}_{\left(r+M-1\right)\text{mod}M,n} & = & \mathcal{S}_{-\left[2w-\left(M-1\right)l\right]}\left\{ \mathcal{S}_{2w-\left(M-1\right)l}\left\{ d_{\textrm{temp}}\right\} \right\} \nonumber \\
\label{eq:integer_signed_disentanglement_partial_M_general}
\end{eqnarray}

\begin{equation}
\hat{d}_{r,n}=\mathcal{S}_{-\left(M-1\right)l}\left\{ \left(-1\right)^{M}\left(d_{\textrm{temp}}-\hat{d}_{M-1,n}\right)\right\} .\label{eq:d_hat_0,n}
\end{equation}
The other outputs can now be disentangled by ($1\leq m<M-2$): 

\begin{eqnarray}
\forall m:\;\hat{d}_{\left(r+m\right)\text{mod}M,n} & = & \mathrm{\mathbf{\delta}}_{\left(r+m\right)\text{mod}M,n}-\mathcal{S}_{l}\left\{ \hat{d}_{\left(r+m-1\right)\text{mod}M,n}\right\} .\nonumber \\
\label{eq:integer_disentanglement_M_general-1}
\end{eqnarray}
Given that for every output position $n$ we are able to recover \emph{all
results of all $M$ streams without using} $\mathrm{\mathbf{\delta}}_{r,n}$
in \eqref{eq:integer_disentanglement_partial_M_general}--\eqref{eq:integer_disentanglement_M_general-1},
the proposed method is able to recover from a single fail-stop failure
in one of the $M$ entangled streams. 

\emph{Remark 4 (dynamic range of generalized entanglement and equivalence
to checksum methods):} Examples for the maximum bitwidth achievable
for different cases of $M$ are given in Table \ref{tab:Examples_l_k}
assuming a 32-bit representation. We also present the dynamic range
permitted by the equivalent checksum-based method {[}\eqref{eq:P_redundant_data_streams}--\eqref{eq:operation_on_inputs-1-1}{]}
in order to ensure that its checksum stream does not overflow under
a 32-bit representation. Evidently, for $M\leq10$, the proposed approach
incurs loss of $1$ to $9$ bits of dynamic range against the checksum-based
method, while it allows for higher dynamic range than the checksum-based
method for $M\geq11$. At the same time, our proposal does not require
the overhead of applying the LSB operations to an additional stream,
as it ``overlays'' the information of each input onto another input
via the numerical entanglement of pairs of inputs. Beyond this important
different, \uline{our approach offers the exact equivalent to checksum
methods of \mbox{\eqref{eq:P_redundant_data_streams}}--\mbox{\eqref{eq:operation_on_inputs_and_redundant_inputs}}
for integer inputs}. Therefore, equivalently to checksum methods,
beyond recovery from single fail-stop failures, our proposal can also
be used for the detection of silent data corruptions (SDCs) in any
input stream, as long as such SDCs do not occur in coinciding output
stream positions. We plan to explore this aspect in future work.

\begin{table}[tbh]
\protect\caption{Examples of $l$ and $k$ values and bitwidth supported for the output
data under $w=32$ bits and: \emph{(i)} different numbers of entanglements;
\emph{(ii)} checksum-based method of \eqref{eq:P_redundant_data_streams}--\eqref{eq:operation_on_inputs-1-1}.
Any failure in 1 out of $M$ streams is guaranteed to be recoverable
under both frameworks. \label{tab:Examples_l_k}}

\centering{}%
\begin{tabular}{|c|c|c|c|c|}
\hline 
\multirow{3}{*}{$M$} & \multirow{3}{*}{$l$} & \multirow{3}{*}{$k$} & \multicolumn{2}{c|}{Maximum bitwidth supported by}\tabularnewline
 &  &  & Proposed: & Checksum-based\tabularnewline
 &  &  & $\left(M-2\right)l+k$ & $w-\left\lceil \log_{2}M\right\rceil $\tabularnewline
\hline 
3 & 11 & 10 & 21 & 30\tabularnewline
\hline 
4 & 8 & 8 & 24 & 30\tabularnewline
\hline 
5 & 7 & 4 & 25 & 29\tabularnewline
\hline 
8 & 4 & 4 & 28 & 29\tabularnewline
\hline 
11 & 3 & 2 & 29 & 28\tabularnewline
\hline 
16 & 2 & 2 & 30 & 28\tabularnewline
\hline 
32 & 1 & 1 & 31 & 27\tabularnewline
\hline 
\end{tabular}
\end{table}

\section{Complexity in LSB Operations with Numerical Entanglement\label{sec:Linear_processing}}

Consider $M$ input integer data streams, each comprising several
samples and consider that an LSB operation $\text{op}$ with kernel
$\mathbf{g}$ is applied on each stream. The operations count (additions/multiplications)
for stream-by-stream sum-of-products between a matrix comprising $M$
subblocks of $N\times N$ integers and a matrix kernel comprising
$N\times N$ integers (see \cite{anastasia2012throughput,goto2008anatomy,murray2008spread,chen2005fault}
for example instantiations) is: $\mathrm{C}_{\text{GEMM}}=MN^{3}$.
For sesquilinear operations like convolution and cross-correlation
of $M$ input integer data streams (each comprising $N$ samples)
with kernel $\mathbf{g}$ {[}see Fig. \ref{fig:Entanglement-with-M_equal_3}(a){]},
depending on the utilized realization, the number of operations can
range from $O\left(MN^{2}\right)$ for direct algorithms (e.g., time-domain
convolution) to $O\left(MN\log_{2}N\right)$ for fast algorithms (e.g.,
FFT-based convolution) \cite{anam2012throughput}. For example, for
convolution or cross-correlation under these settings and an overlap-save
realization for consecutive block processing, the number of operations
(additions/multiplications) is \cite{anam2012throughput}: $\mathrm{C}_{\text{conv,time}}=4MN^{2}$
for time domain processing and $\mathrm{C}_{\text{conv,freq}}=M\left[\left(45N+15\right)\log_{2}\left(3N+1\right)+3N+1\right]$
for frequency-domain processing. 

As described in Section \ref{sec:From-Numerical-Packing-to-Entanglement},
numerical entanglement of $M$ input integer data streams (of $N$
samples each) requires $O\left(MN\right)$ operations for the entanglement,
extraction and recovery per output sample. For example, ignoring all
arithmetic-shifting operations (which take a negligible amount of
time), based on the description of Section \ref{sec:From-Numerical-Packing-to-Entanglement}
the upper bound of the operations for numerical entanglement, extraction
and  recovery is: $C_{\text{ne,conv}}=2MN$. Similarly as before,
for the special case of the GEMM operation using $M$ subblocks of
$N\times N$ integers, the upper bound of the overhead of numerical
entanglement of all inputs is: $C_{\text{ne,GEMM}}=2MN^{2}$. For
all values for $N$ and $M$ of practical relevance (e.g., $100\leq N\leq1000$
and $3\leq M\leq32$) and sesquilinear operations like matrix products,
convolution and cross-correlation, it can easily be calculated from
the ratios $\frac{\mathrm{C}_{\text{ne,GEMM}}}{\mathrm{C}_{\text{GEMM}}}$,
$\frac{\mathrm{C}_{\text{ne,conv}}}{\mathrm{C}_{\text{conv,time}}}$
and $\frac{\mathrm{C}_{\text{ne,conv}}}{\mathrm{C}_{\text{conv,freq}}}$
that the relative overhead of numerical entanglement, extraction and
recovery in terms of arithmetic operations is below $0.3\%$. Most
importantly, 
\begin{equation}
\lim_{N\rightarrow\infty}\frac{\mathrm{C}_{\text{ne,GEMM}}}{\mathrm{C}_{\text{GEMM}}}=\lim_{N\rightarrow\infty}\frac{\mathrm{C}_{\text{ne,conv}}}{\mathrm{C}_{\text{conv,time}}}=\lim_{N\rightarrow\infty}\frac{\mathrm{C}_{\text{ne,conv}}}{\mathrm{C}_{\text{conv,freq}}}=0,
\end{equation}
i.e., the relative overhead of the proposed approach approaches $0\%$
as the dimension of the LSB processing increases. 

On the other hand, the overhead of checksum-based methods in terms
of operations count (additions/multiplications) for each case is represented
by $\mathrm{C}_{\text{cs,GEMM}}=2MN^{2}+\frac{1}{M}\mathrm{C}_{\text{GEMM}}$,
$\mathrm{C}_{\text{cs,conv,time}}=2MN+\frac{1}{M}\mathrm{C}_{\text{conv,time}}$
and $\mathrm{C}_{\text{cs,conv,freq}}=2MN+\frac{1}{M}\mathrm{C}_{\text{conv,freq}}$.
As expected, the relative overhead of checksum methods converges to
$\frac{1}{M}\times100\%$ as the dimension of the LSB processing operations
increases, i.e., 
\begin{eqnarray}
\lim_{N\rightarrow\infty}\frac{\mathrm{C}_{\text{cs,GEMM}}}{\mathrm{C}_{\text{GEMM}}} & = & \lim_{N\rightarrow\infty}\frac{\mathrm{C}_{\text{cs,conv,time}}}{\mathrm{C}_{\text{conv,time}}}\\
 & = & \lim_{N\rightarrow\infty}\frac{\mathrm{C}_{\text{cs,conv,freq}}}{\mathrm{C}_{\text{conv,freq}}}\\
 & = & \frac{1}{M}.\nonumber 
\end{eqnarray}
Therefore, the checksum-based method for fail-stop mitigation leads
to substantial overhead (above $10\%$) when high reliability is pursued,
i.e., when $M\leq8$. Finally, even for the low reliability regime
(i.e., when $M>8$), checksum-based methods will incur more than $4\%$
overhead in terms of arithmetic operations.

\section{Experimental Validation\label{sec:Experiments}}

\begin{figure*}[tbh]
\begin{centering}
\subfigure[$M=3$]{\includegraphics[scale=0.36]{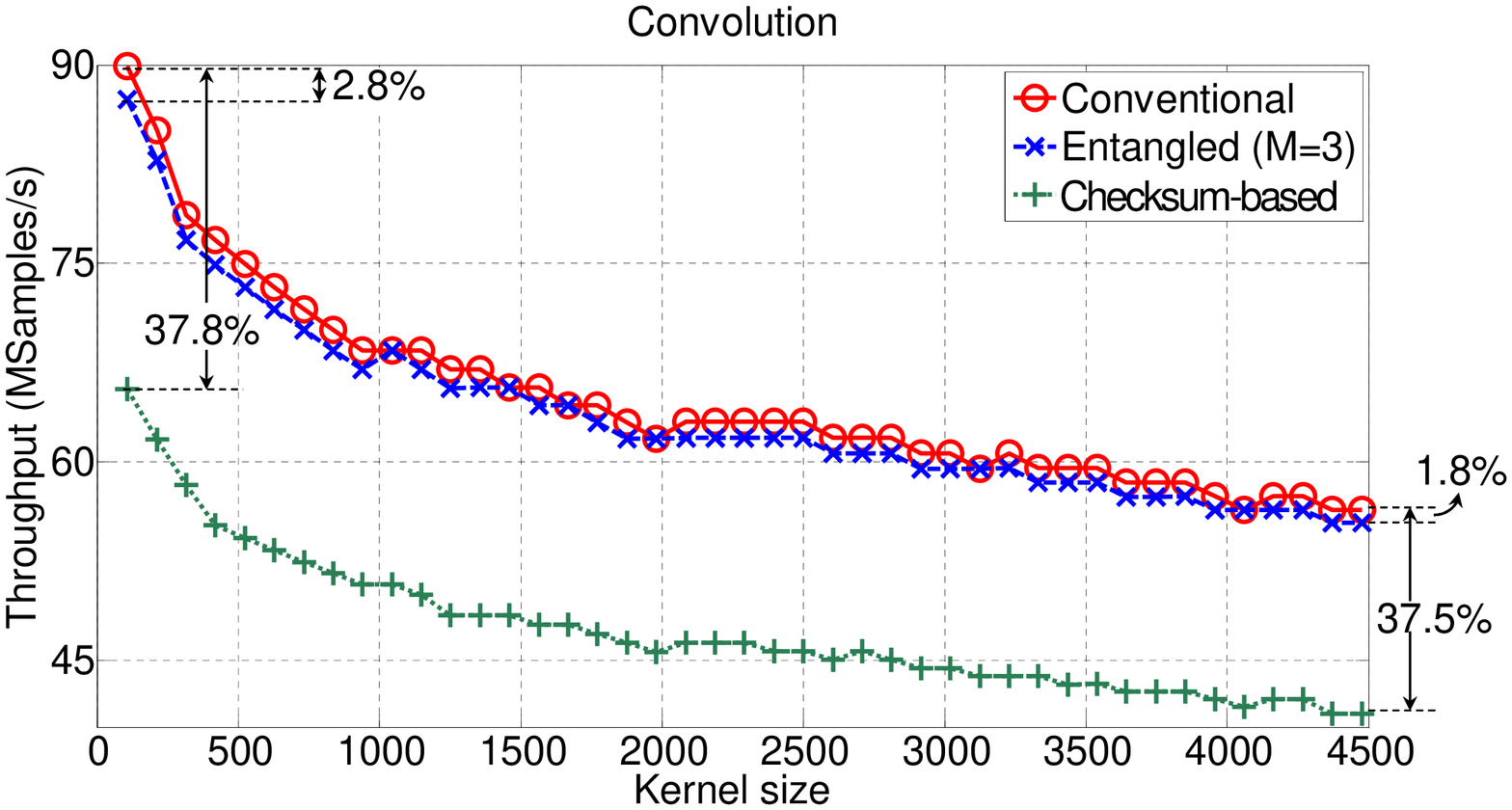}}
\subfigure[$M=8$]{\includegraphics[scale=0.36]{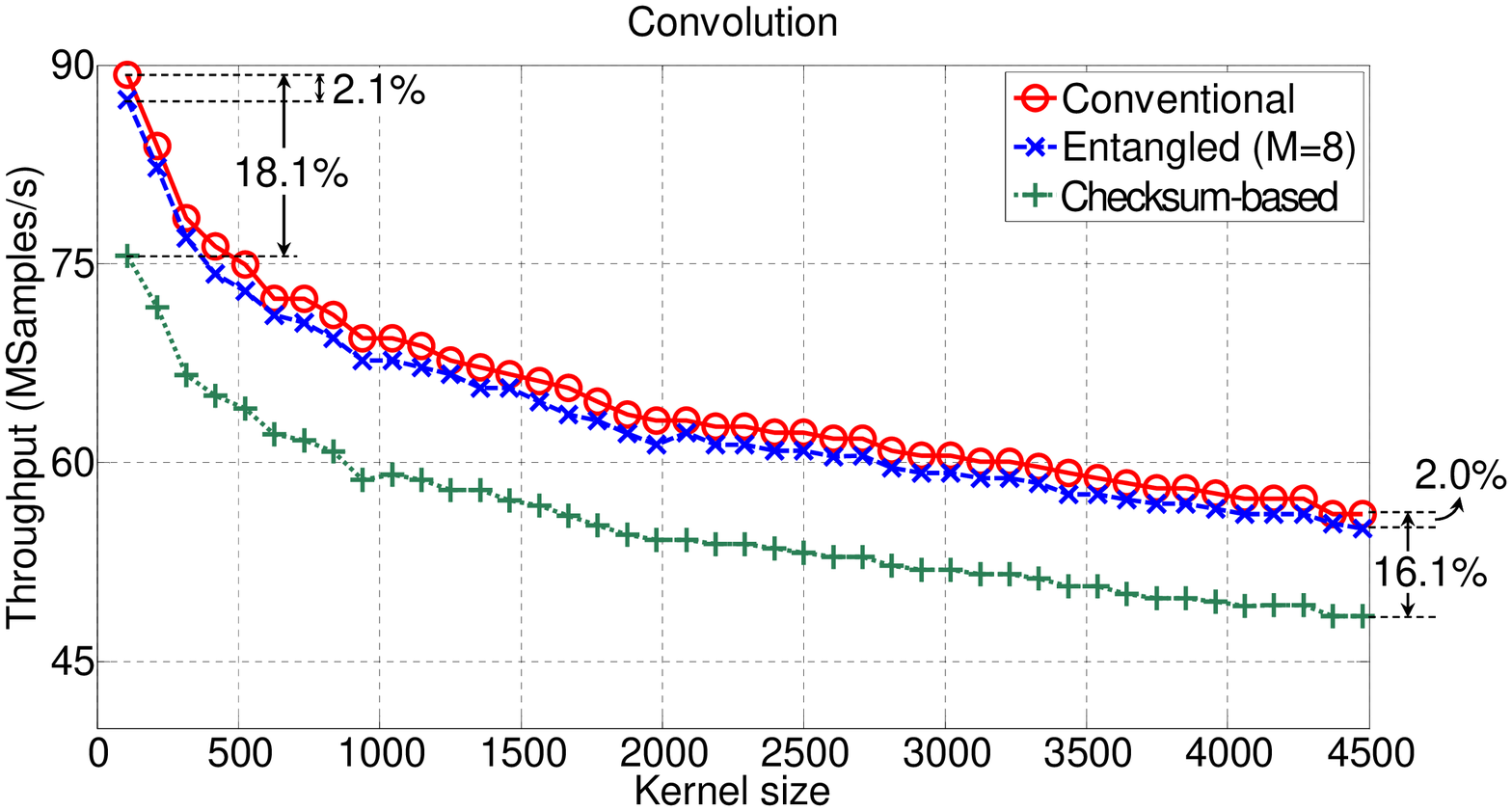}}
\par\end{centering}

\protect\caption{Throughout results for convolution of $M$ integer streams. ``Conventional''
refers to conventional (failure-intolerant) convolution realization
using the state-of-the-art Intel IPP 7.0 library and it is used as
a benchmark under: (a) $M=3$; (b) $M=8$. \label{fig:CONV-benchmark}}
\end{figure*}

All our results were obtained using an Intel Core i7-4700MQ 2.40GHz
processor (Haswell architecture with AVX2 support, Windows 8 64-bit
system, Microsoft Visual Studio 2013 compiler). Entanglement, disentanglement
and fail-stop recovery mechanisms were realized using the Intel AVX2
SIMD instruction set for faster processing. For all cases, we also
present comparisons with checksum-based recovery, the checksum elements
of which were also generated using AVX2 SIMD instructions.

We consider the case of convolution operations of integer streams.
We used Intel's Integrated Performance Primitives (IPP) 7.0 \cite{taylor2003intel}
convolution routine \texttt{ippsConv\_64f} that can handle the dynamic
range required under convolutions with 32-bit integer inputs. We experimented
with: input size of $N_{\text{in}}=10^{6}$ samples, several kernel
sizes between $N_{\text{kernel}}\in\left[100,\:4500\right]$ samples.
Representative results are given in Fig. \ref{fig:CONV-benchmark}
under two settings for the number of input streams, $M$, and without
the occurrence of failures, i.e., when operating under normal conditions%
\footnote{Under the occurrence of one fail-stop failure, the performance of
the proposed approach remains the same as the results are disentangled
as soon as (any) $M-1$ output streams become available. On the other
hand, the performance of the checksum-based approach will decrease
slightly under a fail-stop failure, since results will need to be
recovered from the checksum stream.%
}. The results demonstrate that the proposed approach incurs substantially
smaller overhead for a single fail-stop mitigation in comparison to
the checksum-based method. Specifically, the decrease in throughput
for the proposed approach in comparison to the failure-intolerant
case is only $1.8\%$ to $2.8\%$, while checksum-based method incurs
$16.1\%$ to $37.8\%$ throughput loss for the same test. As expected
by the theoretical calculations of Section \ref{sec:Linear_processing},
this is an order-of-magnitude higher than the overhead of numerical
entanglement.

\section{Conclusions\label{sec:Conclusions}}

We propose a new approach to fail-stop failure recovery in linear,
sesquilinear and bijective (LSB) processing of integer data streams
that is based on the novel concept of numerical entanglement. Under
$M$ input streams ($M\geq3$), the proposed approach provides for:
\emph{(i)} guaranteed recovery from any single fail-stop failure;
\emph{(ii)} complexity overhead that depends only on $M$ and not
on the complexity of the performed LSB operations, thus, quickly becoming
negligible as the complexity of the LSB operations increases. These
two features demonstrate that the proposed solution forms a \emph{third
family} of recovery from fail-stop failures (i.e., beyond the well-known
and widely-used checksum-based methods and modular redundancy) and
offers unique advantages. As such, it is envisaged that it will find
usage in a multitude of systems that require enhanced reliability
against core failures in hardware with very low implementation overhead. 

\bibliographystyle{IEEEbib}
\bibliography{literatur}

\end{document}